\documentclass{article}

\usepackage[english]{babel}

\usepackage[a4paper,top=2cm,bottom=2cm,left=3cm,right=3cm,marginparwidth=1.75cm]{geometry}
\usepackage{float}
\usepackage{caption}
\usepackage{subcaption}
\usepackage{amsmath}
\usepackage{graphicx}
\usepackage[colorlinks=true, allcolors=blue]{hyperref}

\title{Comparative study in fair division algorithms}
\author{Liad Nagi and Moriya Elgrabli \\{\small Instructor: Erel Segal Halevi}}

\begin{document}
\maketitle

\begin{abstract}
\centering A comparison of four fair division algorithms performed on real data from the spliddit website. The comparison was made on the sum of agent's utilities, and the minimum utility for an agent in an allocation.\end{abstract}

\section{Introduction}
A fair division algorithm is an algorithm that divides a set of resources among several people who have an entitlement to them so that each person receives their due share. 
The algorithm takes into account the utilities of the items to each person.
This paper compares fair division algorithms in minimum utility and the sum of all agent's utilities.

\section{Method}
We implemented the three-quarters MMS allocation algorithm from the "An Improved Approximation Algorithm for Maximin Share" article by Jugal Garg and Setareh Taki \cite{approxMMS}, into the fairpy open-source library in python  \footnote{\href{https://github.com/erelsgl/fairpy/blob/master/fairpy/items/approximation_maximin_share.py}{https://github.com/erelsgl/fairpy/blob/master/fairpy/items/approximation\_maximin\_share.py}}.
We compared the performance of this algorithm to three other fair division algorithms implemented in fairpy: Max sum allocation, leximin \cite{wilson1998fair}, and PROPm \cite{baklanov2021propm}.
\newline
We conducted a comparison of 730 cases of item divisions. The data was collected from the spliddit website \cite{goldman2015spliddit} and was kindly shared with us by Nisarg Shah. The data contains 730 cases of items and their evaluations by various agents.
\newline
\newline
We compared the algorithms in two ways: 
\begin{itemize}
\item The minimum utility an agent receives
\item The sum of the utilities of all agents 
\end{itemize}
The comparison was made as an average of the results of each number of agents separately. The average amount of utilities was calculated for one agent, two agents, and so on. Similarly, the minimum utility of an agent in an allocation was calculated separately for each number of agents.  
\newline
\newline
Since the three-quarters MMS allocation algorithm stops when each agent gets at least
three-quarters of his MMS value, it does not divide all items. This can cause a significant gap in performance measurement, so in order to solve this, we added a division of the remaining items as follows: start from the last agent and give him an item he values as more than 0 (if any) and continue dividing the items to other agents until the items run out. We started from the last agent because the Three-quarters MMS allocation algorithm takes care of the first agents first \footnote{\href{https://github.com/erelsgl/fairpy/blob/master/experiments/compare_algorithms.py}{https://github.com/erelsgl/fairpy/blob/master/experiments/compare\_algorithms.py}}.
\newpage
\section{Comparative study}

\begin{figure}[!h]

\begin{subfigure}{.5\textwidth}
\centering
\includegraphics[width=1\linewidth]{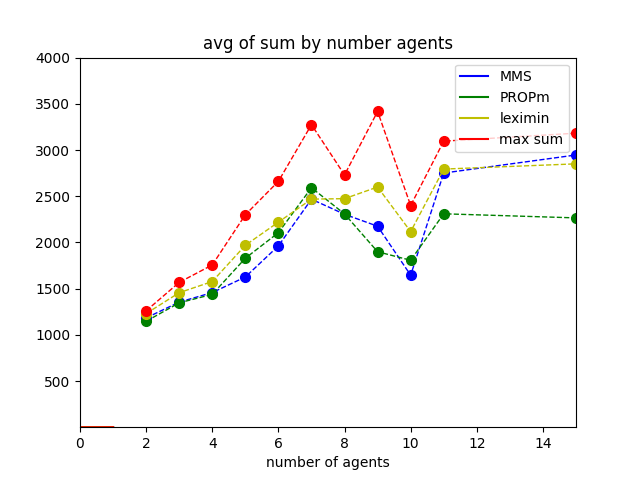}
\caption{\centering The average sum of agents utilities, for each amount of agents} 
\end{subfigure}%
\begin{subfigure}{.5\textwidth}
\centering
\includegraphics[width=1\linewidth]{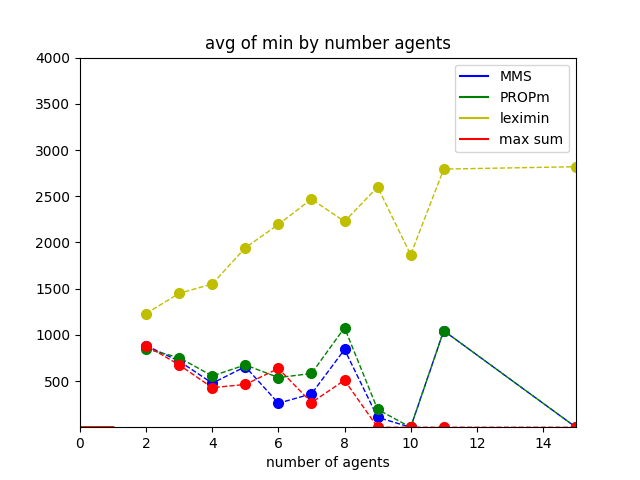}
\caption{\centering The average of the minimum utility for an agent per allocation, for each amount of agents} 
\end{subfigure}%
\newline
\begin{subfigure}{.5\textwidth}
\centering
\includegraphics[width=1\linewidth]{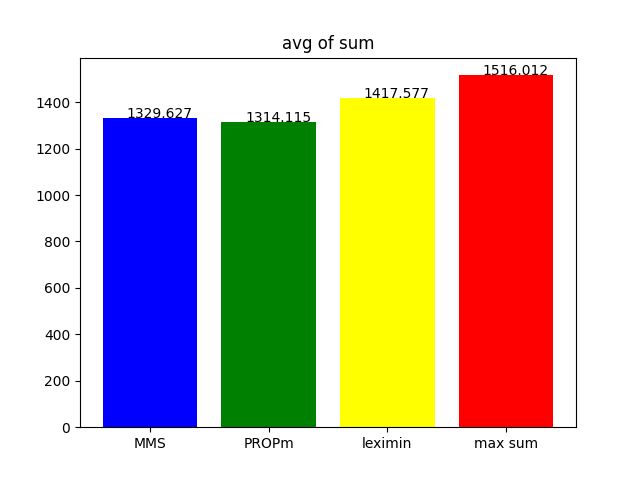}
\caption{\centering The average sum of agents utilities} 
\end{subfigure}%
\begin{subfigure}{.5\textwidth}
\centering
\includegraphics[width=1\linewidth]{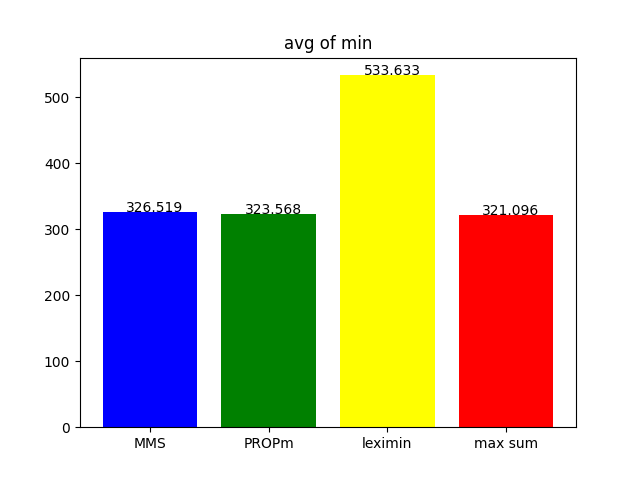}
\caption{\centering The average of the minimum utility for an agent per allocation}

\end{subfigure}

\end{figure}



\section{Results}
As might be expected, for the sum of utilities, the algorithm that maximizes the sum of the utilities returns a significantly better result, and the rest of the algorithms are close in terms of performance.
\newline
In the same way, for the minimum utilities, leximin is significantly better in terms of performance - which makes sense, since leximin allows the division of items in the middle, and therefore the final utility each agent receives can be higher (e.g. if there are two people interested in only one item, both will receive a certain utility), while if items are distributed in their entirety - the minimum utility will be significantly harmed.
\section{Conclusions}
 PROPm and the three-quarters MMS allocation algorithms are close in performance. Although both are based on different calculation methods and try to ensure different fairness conditions, in the end, both return very similar results.
\section {Acknowledgments}
We would like to thank Jugal Garg and Setareh Taki, the authors of the article we implemented, who helped us understanding the article. In addition, we would like to thank Nisarg Shah who shared with us and allowed us to use real data from the spliddit website to perform the experiments on.
\bibliographystyle{alpha}

\bibliography{sample}

\end{document}